\documentclass{vgtc}                          

\ifpdf
  \pdfoutput=1\relax                   
  \usepackage{graphicx}                
  \DeclareGraphicsExtensions{.pdf,.png,.jpg,.jpeg} 
\else
  \usepackage{graphicx}                
  \DeclareGraphicsExtensions{.eps}     
\fi%

\graphicspath{{figures/}{pictures/}{images/}{./}} 

\usepackage{microtype}                 
\PassOptionsToPackage{warn}{textcomp}  
\usepackage{textcomp}                  
\usepackage{mathptmx}                  
\usepackage{times}                     
\usepackage{cite}                      
\usepackage{tabu}                      
\usepackage{booktabs}                  

\usepackage[table]{xcolor} 
\usepackage{tabularx} 
\usepackage[margin=1in]{geometry} 

\definecolor{rowgray}{gray}{0.925}

\onlineid{0}

\vgtccategory{Research}

\preprinttext{Submitted to IEEE VIS Posters, 2024.}

\title{Design Contradictions: Help or Hindrance?}

\author{Aron E. Owen\thanks{e-mail: aron.e.owen@bangor.ac.uk}\\ %
        \scriptsize Bangor University %
\and Jonathan C. Roberts\thanks{e-mail: j.c.roberts@bangor.ac.uk}\\ %
     \scriptsize Bangor University %
}

\authorfooter{
	\item J.C. Roberts, A.E.Owen and P.D. Ritsos are with the School of Computer Science and Electronic Engineering, Bangor University, UK. 
	E-mail: {j.c.roberts; aron.e.owen;}@bangor.ac.uk.}
 
\abstract{

The need for innovative ideas in data visualisation drives us to explore new creative approaches. Combining two or more creative words, particularly those that contradict each other, can positively impact the creative process, sparking novel ideas and designs. As we move towards AI-driven design, an open question arises: do these design contradictions work positively with AI tools? Currently, the answer is no. AI systems, like large language models (LLMs), rely on algorithms that engender similarity, whereas creativity often requires divergence and novelty. This poster initiates a conversation on how to drive AI systems to be more creative and generate new ideas. This research invites us to reconsider traditional design methods and explore new approaches in an AI-driven world. Can we apply the same techniques used in traditional design, like the double diamond model, or do we need new methods for design engineering? How can we quickly design visualisations and craft new ideas with generative AI? This paper seeks to start this critical conversation and offers practical insights into the potential of AI in driving creativity in data visualisation.

} 

\CCScatlist{
  \CCScatTwelve{Human-centered computing}{Visu\-al\-iza\-tion}{Visualization theory, concepts, and paradigms};
}

\begin{document}
\firstsection{Introduction}

\maketitle

The quest for innovation in data visualisation requires us to seek out and embrace new creative approaches continually. This process can positively impact creativity, leading to the development of novel ideas and designs. However, as we transition towards AI-driven design, it is essential to question whether these design contradictions can work positively with AI tools. AI systems, including large language models (LLMs) like OpenAI's GPT series, often need to catch up in this area. These systems rely heavily on algorithms that promote similarity and repetition, whereas true creativity thrives on divergence and novelty. This poster aims to initiate a crucial conversation about how we can drive AI systems to be more creative and generate new, innovative ideas.
We explore a specific example, such as "Round Bar Chart", demonstrating how these contradictions can inspire creativity and lead to suboptimal outputs when handled by generative AI. These examples highlight the need for better strategies to harness AI tools' creative potential.

As we delve deeper into this subject, we are compelled to reassess traditional design methods and contemplate new approaches tailored to the era of AI-driven design. This prompts the question: can established techniques like the double diamond model be effectively applied, or do we need entirely new methods for design engineering? This research underscores the pressing need for the development of new design methods that can effectively harness generative AI to expedite the design of visualisations and foster the creation of new ideas, a pivotal step in the evolution of data visualisation. By understanding and addressing the limitations and possibilities of AI-driven design, we can pave the way for more innovative and effective visualisation solutions, a prospect that holds immense promise for the future of data visualisation.

\section{Related Work and Concepts}
Understanding design contradictions is a collaborative endeavour, requiring the collective insights of researchers from various disciplines. Moreover, this is particularly crucial in generative AI systems, where design contradictions play a significant role. Multidisciplinary research is necessary and a testament to our academic community's collective intelligence and shared responsibility. When handling contradictions in Natural Language Processing (NLP), we can turn to the Stanford Typed Dependencies Representation and word embeddings for contradiction detection, which are not just theoretical constructs but practical tools that have revolutionised our understanding of syntactic and semantic relations within the text. These tools have been pivotal in identifying and managing contradictions within textual data, thereby enhancing NLP models' reliability and real-world applications~\cite{de-marneffe-manning-2008-stanford, Li2017}. Another aspect of contradictions is with the data itself, and there is a comprehensive guide to creating compelling visualisations, including techniques for handling conflicting data~\cite{munzner_2014}, which demonstrates how contradictions may not solely affect generative AI but the data itself can influence. Furthermore, this work emphasises the importance of clear design principles and methodologies to avoid misinterpretations caused by data conflicts. Moreover, this is further explored when discussing various visualisation methods suitable for different data types, including those with inherent conflicts~\cite{heer_bostock_ogievetsky_2010}.

The concept of cognitive load theory, which examines the effects of cognitive load on learning and problem-solving~\cite{sweller_1988}, is crucial for understanding how users interact with AI-generated content, particularly when faced with contradictions. Extending this understanding in a discussion on storytelling in visualisation, highlighting how narrative techniques can reduce cognitive load and enhance user comprehension of complex data~\cite{kosara_mackinlay_2013}. We must also be aware of sound design principles like Tufte's seminal work, which outlines fundamental principles for creating clear and compelling visualisations, emphasising eliminating unnecessary complexity~\cite{tufte_2001}. Building on these principles~\cite{few_2009} offers practical advice on visualisation to accurately and efficiently convey quantitative information. These principles are particularly relevant when addressing the design contradictions posed by generative AI outputs. The literature highlights significant efforts in understanding and addressing design contradictions in generative AI and visualisation design. The studies underscore the importance of clear design principles, effective contradiction detection, and the reduction of cognitive load to enhance user satisfaction. Ongoing interdisciplinary research is essential to develop more reliable AI models capable of handling these contradictions, ultimately improving AI-generated content's utility.

\begin{figure}[th]
\includegraphics[width=0.8\columnwidth]{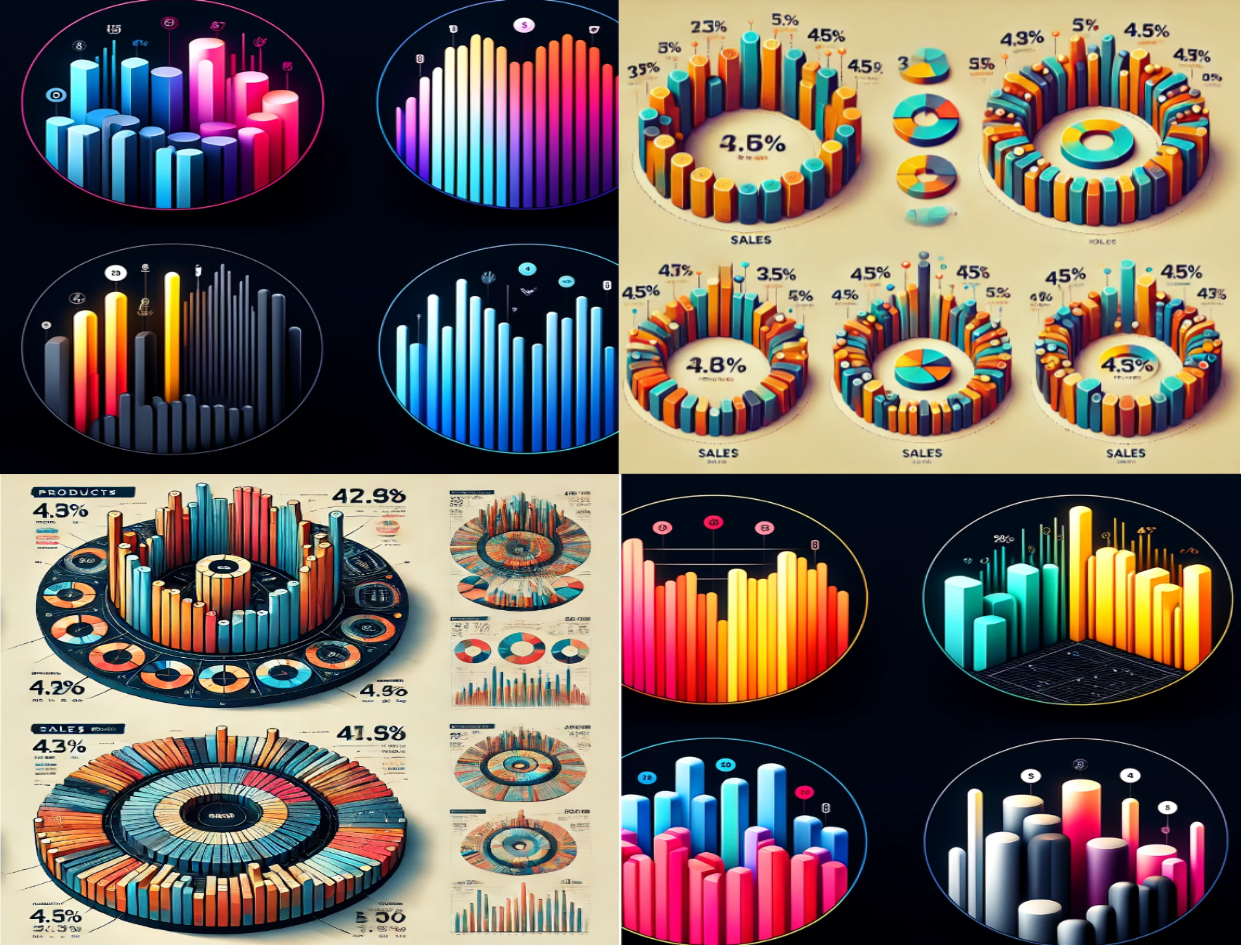}
\centering
\caption{\label{fig:gptbar}This image demonstrates the design contradiction of round bar charts generated by ChatGPT for visualizing sales data. Each chart explores different styles, highlighting the challenge of maintaining clarity and comparability in a circular format.\vspace{-5mm}}

\end{figure}

\section{Case Study - Round Bar Chart}
Designers, always on the lookout for fresh and unconventional sources of inspiration, may find a spark in the world of sound engineering, Specifically, the circular visualisers used in audio applications. Furthermore, this intriguing concept raises a question: Can we apply the circular effect to the data of a traditional bar chart, creating a unique and engaging visualisation? A round bar chart is inherently contradictory because bar charts are designed to use rectangular bars to compare categorical data through their lengths. Making them round contradicts the primary purpose of the bar chart, which is to show differences in length clearly. However, let us investigate this use case further. Imagine attempting to represent sales data for different products using a round bar chart. This design, with its circular shape and bars radiating outwards, is undeniably eye-catching. Yet, it presents a host of challenges. The radial arrangement makes it difficult to compare bar lengths, and the potential for overlapping or converging bars adds visual clutter. The chart may also include unnecessary design elements that distract from the data, further complicating the visualisation.

Further analysis of the output, as seen in Fig~\ref{fig:gptbar}, demonstrates that the round bar chart needs to achieve the clarity of a traditional bar chart. The intended purpose of comparing categorical data through bar lengths is compromised. This analysis represents the hindrance aspect of our design contradiction.
Despite the challenges, the round bar chart offers unique and exciting possibilities. Its circular design can be a visual feast, capturing attention in presentations, reports, or dashboards. It has a contemporary and innovative appearance, making it a perfect fit for design-forward or creative industries. It provides a compact representation that can fit into confined spaces, such as a corner of a dashboard or a compact infographic. A round bar chart can intuitively represent this cyclic nature, emphasising continuity and periodicity for data with a natural cyclic pattern (e.g., hours in a day, months in a year). The unique format can be used creatively to tell a story or present data in a novel way, engaging the audience more deeply than traditional charts. The distinctive visual form can make the data more memorable and impactful.
While round bar charts introduce challenges in data interpretation, they offer significant visual and aesthetic benefits. They can provide compelling data representations that capture attention, highlight cyclical patterns, and enhance storytelling when used thoughtfully. The key is to balance these visual benefits with the need for clarity and accuracy in data representation.
This case study asks: Do design contradictions help or hinder the creative process in visualisation? Can we leverage these contradictions to push the boundaries of design while maintaining the integrity of the data presented? This exploration invites further discussion on the potential of AI-driven creativity and the development of new design methodologies in data visualisation.

\begin{figure}[th]
\includegraphics[width=0.8\columnwidth]{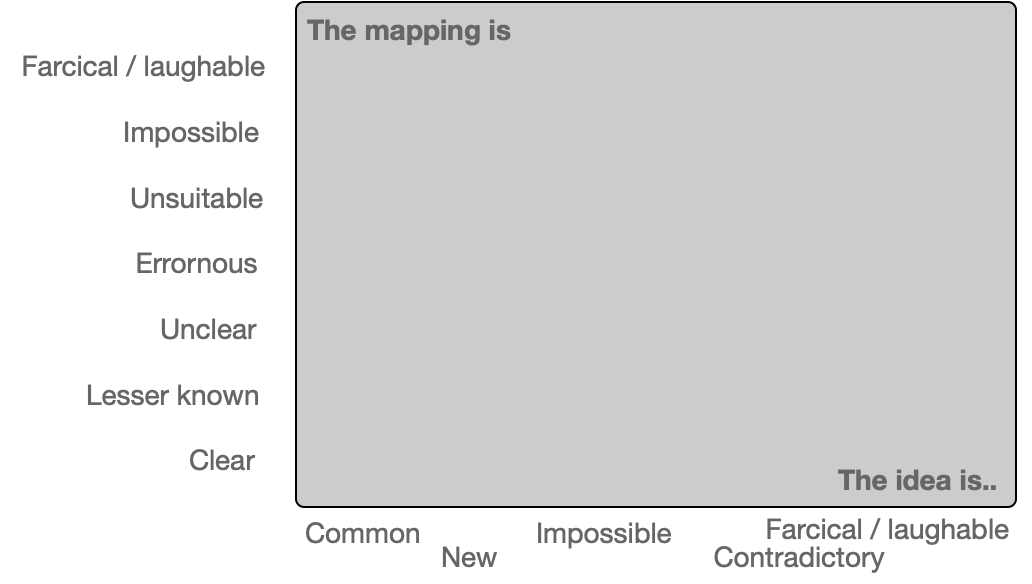}
\centering
\caption{\label{fig:mapping}This diagram maps the dimensions of design contradictions in visualisation. The vertical axis (“The mapping is…”) ranges from clear to farcical, while the horizontal axis (“The idea is…”) spans from common to contradictory. It helps analyse the balance between innovative ideas and practical execution in visualisations.\vspace{-5mm}}

\end{figure}

\section{Conclusion}

This paper ventures into a unique and unexplored area of research, delving into the design contradictions that large language models (LLMs) such as OpenAI's GPT. These 'design contradictions' are the inherent conflicts that arise when LLMs attempt to visualize data with contradictory elements like round bar charts. The models often resort to visual shortcuts that compromise the clarity and accuracy of the output, underscoring the need for more sophisticated handling of such contradictions. Our example vividly illustrates how LLMs grapple with these contradictions, often leading to cluttered and unclear outputs, such as bar charts with overlapping bars. These examples underscore the significant challenges in maintaining data integrity and clarity when faced with contradictory design requirements.

We warmly invite the research community to join us on this journey of continuous improvement. By working together, we can develop more robust LLMs capable of effectively handling design contradictions.

\bibliographystyle{abbrv-doi}

\bibliography{template}
\end{document}